\documentstyle[aps,graphicx,preprint,tighten]{revtex}

\newcommand{\be}{\begin{equation}}
\newcommand{\ee}{\end{equation}}
\newcommand{\bea}{\begin{eqnarray}}
\newcommand{\eea}{\end{eqnarray}}
\newcommand{\nn}{\nonumber}
\newcommand{\qc}{\quad ,}
\newcommand{\qp}{\quad .}
\newcommand{\rmi}{{\rm i}}
\newcommand{\rme}{{\rm e}}

\newcommand{\lapprox}{{\raisebox{-0.5ex}{${\scriptstyle<}$} 
\atop \raisebox{0.5ex}{${\scriptstyle\sim}$}}}

\newcommand{\bsl}[1]{#1 \!\!\! /}
\newcommand{\eq}[1]{Eq.~(\ref{#1})}
\newcommand{\msc}[1]{Sec.~\ref{#1}}
\newcommand{\fig}[1]{Fig.~\ref{#1}}
\newcommand{\grad}{\vec{\nabla}}
\newcommand{\cs}{cross section }
 
\newcommand{\po}{production operator }

\def\Journal#1#2#3#4{{#1} {\bf #2}, (#4) #3}
\def\NP{{\em Nucl. Phys.} }
\def\NC{{\em Nuovo Cim.} }
\def\PL{{\em Phys. Lett.} }
\def\PR{{\em Phys. Rev.} }
\def\PRL{{\em Phys. Rev. Lett.} }

\def\AP{{\em Ann. of Phys.} }
\def\ZP{{\em Z. Phys.} }

\begin{document}
\begin{center}
{\Large \bf

Coherent Photoproduction of Pions on 
{ Spin-Zero} 
Nuclei
in a relativistic, non-local Model
\footnote{Work supported by BMBF and GSI Darmstadt.}
\footnote{This work forms part of the dissertation of W. Peters.}
}
\bigskip
\bigskip

{W. Peters\footnote{Wolfram.Peters@theo.physik.uni-giessen.de}, 
H. Lenske and U. Mosel}

\bigskip
\bigskip
{ \it
Institut f\"ur Theoretische Physik, Universit\"at Giessen\\
D-35392 Giessen, Germany\\ }

\end{center}

\begin{abstract}
  A relativistic and non-local model for the coherent photoproduction
  of pions on { spin-zero} nuclei is presented. The production
  operator is derived from an effective Lagrangian model that contains
  all the degrees of freedom known to contribute to the elementary
  process.  Using the framework of DWIA, the matrix elements of this
  operator are evaluated using relativistic bound state wave
  functions. Final state interactions are accounted for via a
  pion-nucleus optical potential. The effects of in-medium
  modifications of the production operator are discussed. We give
  results for $^{12}{\rm C}$ and $^{40}{\rm Ca}$, and compare our calculations to
  the data of the A2 collaboration at MAMI.
\end{abstract}

\section{Introduction}

Photonuclear reactions offer a unique possibility to test our
understanding of hadrons in vacuum as well as within a nuclear
environment. For elementary reactions like the photoproduction of
pions on nucleons, relativistic models based on an effective
Lagrangian approach are well established by now (see for example
\cite{feuster} and references therein). Starting from the elementary
process, numerous studies treat the photoproduction of pions on
nuclei. Among the large number of possible reactions, the coherent
photoproduction plays a special role. Since in this case the nucleus
remains in the ground state after the reaction process, the amount of
further theoretical assumptions is smaller than e.g. in the case of
reactions like $(\gamma,\pi N)$ or completely inclusive processes like
photoabsorption on nuclei.  In this sense the coherent photoproduction
of pions is the cleanest test for any model that treats
photoproduction of pions on nuclei.

Data on the coherent photoproduction of pions on nuclei were published
in \cite{arends}. The extraction of the coherent events in most of
these experiments, however, was uncertain, so that comparison to these
data is not very conclusive. More recently, the differential cross
section was given for a photon energy of $E_\gamma \approx 175$ MeV in
\cite{gothe}.  It is, however, a special feature of the coherent
process, that the differential cross section is mainly sensitive to
the nuclear form factor (see Sec. \ref{theocoh}). Since the nuclear
wave functions, and therefore the bound state properties of the
nucleus, are taken as an input, all models yield more or less the
correct shape of the differential cross section as can be seen in
\cite{gothe}. For a given energy they differ mainly with respect to
the absolute value of the cross section. To further distinguish
between different models, energy dependent data are needed. Thus the
new data of the A2-collaboration at MAMI \cite{schmitz}, which give
the differential cross section at a fixed angle for a large energy
range between 140 MeV and 430 MeV, are highly welcome.  They enable us
to test our calculations not only close to threshold, but over the
entire $\Delta$ region.

Part of the existing calculations for the coherent photoproduction of
pions work in the delta-hole model
\cite{carrasco,lakti,koerfgen,oset,kara,koch,koch0,saha,kling}.  These
calculations treat the $\Delta$ and pion dynamics in a very
sophisticated manner, however, non-resonant contributions 
are normally neglected (See, however, \cite{oset,koch0,saha}).  
{ 
While many other works use a local approximation, in \cite{saha}
the non-locality of the process is treated explicitly.}  
Another group of models is mainly used for the threshold region and employs
the distorted wave impulse approximation (DWIA)
\cite{giri,boffi1,boffi2,chumba1,chumba2}. These studies take into account
the non-resonant contributions, but resort for technical reasons to
non-relativistic and local approximations.

{
  
  In this paper we present a relativistic, non-local approach to the
  coherent photoproduction of neutral pions on spin-zero nuclei in the
  framework of DWIA.  The production operator is taken from an
  effective field theory, which describes the cross section for the
  production of neutral pions on nucleons well.  The wave function of
  the outgoing pion is calculated using an empirical $\pi$-nucleus
  optical potential including s- and p-wave contributions.  The bound
  state wave functions are extracted from a relativistic equation of
  motion with empirically determined scalar and vector potentials. We
  thus include resonant and non-resonant contributions to
  the production operator in a consistently relativistic manner,
  without non-relativistic or local approximations.  We also study the
  effect of medium modifications of the production operator,
  especially in the $\Delta$ contribution.  This approach enables us
  to test our understanding of the relativistic structure of the
  nuclear ground state and the production operator, though we resort
  to empirical models for the nucleus and the pion-nucleus
  interaction.  }

In the following section we first describe the elementary production
operator we use. After some general remarks on the coherent
photoproduction of pions, we then give the details
of our model. Results for
PWIA, as well as DWIA calculations are shown in \msc{results}. We
finally discuss the effects of in-medium modifications of the 
production operator.
In \msc{sum} the results of our study are summarized.

\section{The model}
\subsection{The elementary operator}
\label{theon}
The diagrams contributing to the elementary process of photoproduction
of neutral pions on a single nucleon are shown in Fig.~\ref{diagrams}:
There are the direct and crossed nucleon graph, the direct and crossed
$\Delta$-graph and the graph containing the omega meson in the
$t$-channel. There could in principle also be an analogous graph
containing the rho meson, but due the smallness of the corresponding
couplings it contributes only negligibly to the production of neutral
pions \cite{feuster}. We use the following interaction terms:

\bea
\label{lint}
{\cal L}_{\pi NN} &=& - \:\frac{g_{_{\pi NN}}}{m_\pi} \: 
                       \overline{\psi} \: \gamma_5
                      (\bsl \partial  \vec{\pi}) \: \vec{\tau} \: \psi
                      \nn\\
                      \nn\\
{\cal L}_{\gamma NN} &=& - {e} \: 
                       \overline{\psi} \: {\textstyle \frac{1}{2}}(1+\tau_3) 
                       \gamma_{\mu} \: 
                      \: \psi \: A ^\mu \nn\\
                      & & - {\textstyle \frac{1}{2}} \:{\rm e} \:
                      \frac{\kappa_p}{ 2 m_{_N}} \:
                      \overline{\psi} \: {\textstyle \frac{1}{2}}(1+\tau_3)\:
                      \sigma_{\mu \nu} \:\psi \:F^{\mu \nu }\nn\\
                      & & - {\textstyle \frac{1}{2}} \:{\rm e} \:
                      \frac{\kappa_n}{ 2 m_{_N}} \:
                      \overline{\psi} \: {\textstyle \frac{1}{2}}(1-\tau_3)\:
                      \sigma_{\mu \nu} \:\psi \:F^{\mu \nu }\nn\\
                      \nn\\
{\cal L}_{\pi N\Delta} &=&  \frac{g_{_{\pi N \Delta}}}{m_\pi} \: 
                       \overline{\psi}_\Delta^\mu \: 
                      ( \partial_\mu  \vec{\pi}) \: \vec{T} \: \psi
                       + {\rm h.c.}\nn\\
                      \nn\\
{\cal L}_{\gamma N\Delta} &=&  \frac{ie g_{_{\gamma N \Delta}}}{m_\pi} \: 
                       \overline{\psi}_\Delta^\mu \: 
                       \gamma^\nu \:\gamma_5\: T_3 \: \psi \: F_{\mu\nu}
                       + {\rm h.c.}\nn\\
                      \nn\\
{\cal L}_{\omega \pi \gamma} &=&  \frac{e g_{_{\omega \pi \gamma}}}{4m_\pi} \: 
                       \varepsilon_{\mu\nu\rho\sigma}  
                       \:(V^{\mu\nu} \: F^{\rho\sigma})\pi\nn\\
                      \nn\\
{\cal L}_{\omega NN} &=& -  g_{\omega NN}^v \: 
                       \overline{\psi} \: 
                       \gamma_{\mu} \: 
                      \: \psi \: \omega ^\mu \nn\\
                      & & - {\textstyle \frac{1}{2}} \:  
                      \frac{g_{\omega NN}^t}{ 2 m_{_N}} \:
                      \overline{\psi} \: 
                      \sigma_{\mu \nu} \:\psi \:V^{\mu \nu }
\qc
\eea
with $ F_{\mu\nu} = \partial_\mu A_\nu - \partial_\nu A_\mu$ and $
V_{\mu\nu} = \partial_\mu \omega_\nu - \partial_\nu \omega_\mu$, where
$A_\mu(\omega_\mu)$ denotes the photon (omega) field.

For the $\Delta$-propagator we take \cite{ew}:

\bea
G_\Delta^{\mu\nu}(p) &=& \rmi \:
 \frac{\bsl p + m_\Delta }
{s-m_\Delta^{2} + \rmi \sqrt{s}\:\Gamma(s)} \:
\Lambda^{\mu\nu}
\label{delprop12}
\eea
with

\bea
\Lambda^{\mu\nu}\:=\:
\left( g^{\mu \nu} -
\frac{1}{3}\gamma^{\mu}\gamma^{\nu} -
\frac{2}{3m_\Delta^{2}}p^{\mu}p^{\nu} - 
\frac{1}{3m_\Delta} 
(\gamma^{\mu} p^{\nu} - p^{\mu} \gamma^{\nu}) \right) 
\qp
\label{delprop32}
\eea
The width of the $\Delta$ is taken to depend on the energy:

\bea
\label{deldec}
\Gamma(s) &=& \Gamma_o \frac{m_\Delta}{\sqrt{s}} 
\left(\frac{q}{q_o}\right)^3 \left( \frac{q_o^2 + c^2}{q^2+c^2} \right)^2
\eea
with $\Gamma_o$ = 120 MeV and $c$ = 0.5 GeV. $q$ and $q_o$ denote the
momentum of the pion resulting from a decay of a $\Delta$ of a mass
$\sqrt{s}$ and of 1.232 GeV, respectively, in the rest-frame of the
$\Delta$.

For the coupling constants we use the following values:

\be
\label{couplcons}
\begin{array}{lclclcl}
e &=& 0.3028 &,&  g_{_{\pi NN}} &=& 0.97  \\
\kappa_p &=& 1.79 &,&  \kappa_n &=& -1.91  \\
g_{_{\pi N \Delta}} &=& 2.1 &,&  g_{_{\gamma N \Delta}} &=& 0.337  \\
g_{_{\omega \pi \gamma}} &=& 0.313 &,&  g_{_{\omega NN}}^v &=& 10  \\
g_{_{\omega NN}}^t &=& 1.4 
\qp
\end{array}
\label{couplings}
\ee

The coupling constants of the photon and the pion to the nucleon are
well known. The value we use for $g_{_{\pi N \Delta}}$ corresponds to
an on-shell decay width of the $\Delta$ of 120 MeV, so that it is
consistent with $\Gamma_o$ in \eq{deldec}. $g_{_{\omega \pi \gamma}}$
has been determined from the decay width of an omega meson into a pion
and a photon \cite{feuster}.  For the couplings of the omega to the
nucleon different values are used in the literature;  the values given
in (\ref{couplings}) are taken from \cite{grein}. The contribution of
the omega graph to the photoproduction of pions or eta mesons is
small, so that the $\omega NN$ coupling constants cannot be determined
by fitting the experimental data for these processes
\cite{mukop,davidson}. Instead, they are taken from $NN$ scattering
\cite{dumbrajs}, or they are derived from the $\rho NN$ parameters via
$SU(3)$ considerations \cite{garci}. The values for $g_{_{\omega
    NN}}^v$ used in the literature for photoproduction processes range
from 8 \cite{garci} to 17 \cite{laget}.  In addition, a form factor is
introduced at the $\omega NN$-vertex:

\bea
\label{omff}
F(t) &=& \frac{\Lambda^2 - m_\omega^2}{\Lambda^2 -t}
\qc
\eea
with $\Lambda^2$ = 2 GeV$^2$ \cite{mukop}. The sensitivity of our
results to the omega parameters will be discussed in Sec.
\ref{results}.

{ Finally, $g_{_{\gamma N \Delta}}$ and $c$ in \eq{deldec} were
  determined by comparing the results of the present model to the data
  for the cross section of the photoproduction of neutral pions on
  protons \cite{pdat}. As can be seen from \fig{gamnuc}, the
  differential cross section can be well reproduced with the
  parameters given above.  We did not use multipoles for the
  determination of these parameters, since multipoles are especially
  sensitive to unitarization effects, which are not included in the
  present model.
}

\subsection{Photoproduction on the nucleus}
\label{theonucl}

After having specified the elementary production operator we now turn
to the photoproduction on the nucleus. The framework commonly used for
exclusive processes like the coherent photoproduction is the DWIA.
Here it is assumed that the production process involves only a single
nucleon, while for the distortions of in- or outgoing particles
interactions with the entire nucleus are taken into account.  {
  Thus many-body contributions to the production process are not taken
  into account.}  The single-particle production operator is taken
from an elementary model and evaluated using { relativistic} bound
state and scattering wave functions instead of plane waves.
Since the nucleon participating in this reaction is bound in a
potential, it is off-shell, and the production operator must be
evaluated for kinematical situations different from the one in the
free case.

For technical reasons a relativistic production operator is often
simplified using on-shell relations \cite{pieka} or a non-relativistic
reduction \cite{laget}, in order to make its evaluation easier.  The
resulting production operators are more or less equivalent on-shell,
but the off-shell behavior is not the same as for the original
operator. To avoid this problem, we take the production operator in
its original form from the model described in the previous section,
without rewriting it. Thus we use the natural off-shell dependence
resulting from an effective field theory.

A numerical complication arises from the fact that the production
operator depends on the momentum of the struck nucleon, i.e. it is
non-local. To circumvent complicated integrations, most DWIA
calculations evaluate the production operator in a local approximation
at some fixed effective nucleon momentum.  For knock-out
processes like $(\gamma,\pi N)$ the asymptotic momenta of the
outgoing particles can be used to estimate the momentum arguments of
the production operator. In the case of coherent production, however,
both the incoming and the outgoing nucleon are in a bound state, so
that they do not have a well defined asymptotic momentum.  The
validity of the local approximation for the photoproduction of charged
pions is is discussed in \cite{li,tiator}.  To avoid the uncertainties
related to the local approximation, we evaluate the matrix elements of
the production operator non-locally. We will discuss the local and the
non-relativistic approximation further in Sec.~\ref{results}.

\subsection{The coherent process}
\label{theocoh} 

In the nuclear coherent photoproduction of pions, the nucleus
remains in the ground state after the reaction. 
Since the pion in the final state has the quantum numbers $0^-$,
the Lorentz invariant amplitude for the
entire process must have the form \cite{pieka}:

\bea
\label{amplfund}
T^{(\lambda)} = \: \varepsilon_\mu^{(\lambda)}\: T^{\mu}
\eea
with
\bea
\label{tmu}
T^{\mu} = \varepsilon^{\mu\nu\rho\sigma} 
                \: k_\nu \:  
                p_\rho \: q_\sigma \:\:
                A(s,t)
\qp
\eea
$k$, $p$ and $q$ are the momenta of the incoming photon, the
incoming nucleus and the outgoing pion, respectively.
$\varepsilon_\nu^{(\lambda)}$ is the polarization vector of the photon
and $A(s,t)$ is a scalar function that contains the entire dynamics of
the process and depends on the Mandelstam variables $s$ and $t$. This
can be shown to yield:

\bea
\label{sin2}
\sum\limits_\lambda\: \mid T^{(\lambda)} \mid^2 = \: W^2 \: 
                     k_{cm}^2 \:  q_{cm}^2
                     {\rm sin}^2 \theta_{cm} \mid A(s,t)\mid^2
\qp
\eea
$k_{cm}^2 $ and $ q_{cm}^2$ are the three momenta of the photon and
the pion, $\theta_{cm}$ is the scattering angle and $W$ is the total
energy of the photon-nucleus system; all these quantities are taken in
the cm-frame.  Thus the well known $\sin^2 \theta$-dependence of the
coherent photoproduction of pions results directly from the quantum
numbers involved.  The differential cross section in the cm-frame is
then given by:

\bea 
\label{cs}
\frac{d\sigma}{d\Omega} \:=\: \left( \frac {M_A}{4 \pi W}\right)^2 \:
                     \frac{q_{cm} }{  k_{cm}}\:\frac{1}{2}\:
                     \sum\limits_\lambda\: \mid T^{(\lambda)} \mid^2
\qc
\eea
where $M_A$ is the mass of the nucleus.

Another important consequence of \eq{amplfund} is, that if one replaces
the photon polarization $\varepsilon_\mu$ by the photon momentum
$k_\mu$, $T^{(\lambda)}$ vanishes.  Thus the amplitude $T^{(\lambda)}$
is gauge invariant from the very beginning, independent of the model
used for the nuclear ground state or the production process. This
is a special feature of the coherent photoproduction
{ on spin-zero nuclei},
which makes this reaction even more attractive from a theoretical
point of view: In other reactions than the coherent one, like
$(\gamma,\pi N)$ or $(e,e' N)$, the usual DWIA approach leads to a
gauge-dependent amplitude \cite{johan,pollock,gil}. Methods to
restore gauge invariance lead to theoretical
uncertainties \cite{pollock}, which do not occur in the case of
coherent photoproduction. 

We work in position space, since the bound state and scattering state
wave functions can easily be obtained in a position space
representation. 
In the approach described above 
the first diagram in
Fig.~\ref{diagrams} corresponds to the following non-local expression:

\bea
\label{amplex}
T_{N dir}^{(\lambda)} =
\sum\limits_{\alpha \:occ.} \: \int d^3 x \: d^3 y \:
      \overline{ \psi}_\alpha(\vec{x}) \: {\phi_{\pi}^{(-)}}^*(\vec{x}) \:
      \Gamma_{\pi NN}\:G_N^o (E;\vec{x},\vec{y})\:\Gamma_{\gamma NN}^\mu\:
      \phi_\mu^{(\lambda)} (\vec{y})\: \psi_\alpha(\vec{y})
\qp
\eea
Here $\psi_\alpha$ is the wave function of the bound nucleon,
$\phi_\mu^{(\lambda)}$ is the wave function of the photon, which can
be assumed to be a plane wave and $\phi_{\pi}^{(-)}$ is the distorted wave
function of the pion satisfying incoming boundary conditions
\cite{joach}.  $\Gamma_{\pi NN}$ and $\Gamma_{\gamma NN}^\mu$ are the
vertices resulting from the coupling terms in \eq{lint}
and $G_N^o$ is the free nucleon propagator:

\bea
\label{nprop}
G_N^o (E;\vec{x},\vec{y}) = \int \: \frac{d^3p}{(2\pi)^3} \:
            \frac{\rmi \rme^{\rmi \vec{p} ( \vec{x}-\vec{y})}}
                 {\bsl p - m}
\qp
\eea
The effect of including a dressed instead of a free nucleon propagator
will be discussed in \msc{mediumeffects}.
$E$ is naturally determined by energy conservation:

\bea
\label{econs}
E = E_\gamma + E_\alpha
\qc
\eea
where $E_\alpha$ is the total, relativistic energy of the bound nucleon.

To evaluate (\ref{amplex}), we use partial-wave expansions for the
wave functions as well as for the propagator 
\cite{joach}. Whenever derivatives appear in the vertices, we actually
insert the derivatives of wave functions, so that we take into account
the 
non-locality of the production operator. 
After inserting the partial-wave expansions into \eq{amplex} we
can perform the angular integrations analytically, since they only
involve spherical harmonics. The two radial integrations are
evaluated numerically. 

The remaining graphs in Fig.~\ref{diagrams} are treated analogously.
For the $\Delta$-propagator we treat the contraction of
$\Lambda_{\mu\nu}$ from \eq{delprop32} with the $\Delta N
\gamma$-vertex as one term and use the partial-wave expansion only for
the remaining spin-$\frac{1}{2}$ part of the propagator:

\bea
 \frac{\bsl p + m_\Delta }
{s-m_\Delta^{2} + \rmi \sqrt{s}\:\Gamma(s)} \:
\qc
\eea
{ by taking its Fourier transform analogously to \eq{nprop}.  However,
  a closed expression of this Fourier transform, which is needed in
  order to use the formula for a partial wave expansion, can only be
  obtained if the imaginary part of the denominator does not depend on
  the three-momentum $\vec{p}$.  Since $s\:\Gamma(s)$ depends on
  $\vec{p}$ via $s=p_o^2 - \vec{p}\:^2$, we evaluate this term at a
  value $s_o(E_\gamma,E_\alpha)$, which is taken to be the invariant
  mass of a system of the photon and the struck nucleon, averaged
  over the Fermi-sphere.  Consequently, the $\Delta$-width is a
  function of the energy of the incoming photon ($E_\gamma$) and the
  struck nucleon ($E_\alpha$) .}

A further technical problem arises from the $p_\mu\:
p_\nu$ term in \eq{delprop32}, since its spatial components $p_i\:p_j$
contain second derivatives, an exact treatment of which would greatly
complicate the calculations. We therefore approximate this term by
replacing

\bea
p_i\:p_j  \:\to\: k_i\:k_j
\qc
\eea
where $k$ is the three-momentum of the photon.  Since in momentum
space the three-momentum of the $\Delta$-propagator is the sum of
photon and nucleon momentum, this amounts to putting for this special
term the three-momentum of the incoming nucleon to zero. Using this
approximation, this term contributes only on the percent level, thus
affecting our results insignificantly.

The $\omega NN$ form factor in \eq{omff} is approximated by using the
asymptotic momentum of the pion in 
$t=(p_\gamma - p_\pi)^2$. In the relevant kinematic region, $t$
is small anyway as compared to the cutoff $\Lambda^2$. Consequently
the dependence of $F(t)$ on the pion distortions is weak and its main
effect is a renormalization of the $\omega NN$ coupling.

The angular integrations in \eq{amplex} lead to rather
complicated expressions for the vertices depending on the incoming
and outgoing angular momenta. The $\sin  \theta$ dependence of the
differential cross section is one test for their correct numerical
implementation. As a further test we have
checked that for $T^\mu$ from \eq{tmu} our calculation yields:

\bea
T_o = 0;\:T_i\: k_i = 0
\qc
\eea
which has to be fulfilled for each graph separately, as can be seen
from  the definition of this quantity.

Even though we evaluate the amplitude in position space it is
instructive to consider \eq{amplex} in momentum space:

\bea
\label{amplexmom}
T_{N dir} &=&
\sum\limits_{\alpha \:occ.}  
\int  \frac{d^3 p}{(2\pi)^3} \: 
      \overline{ \psi}_\alpha(\vec{p}+\vec{k}-\vec{q}) \:
      \Gamma_{\pi NN}\:G_N (E;\vec{p}+\vec{k})\:\Gamma_{\gamma NN}\:
      \psi_\alpha(\vec{p}) \nn \\
 &=& 
\sum\limits_{\alpha \:occ.} \: 
\int  \frac{d^3 p}{(2\pi)^3} \: 
      \overline{ \psi}_\alpha(\vec{p}+\vec{k}-\vec{q}) 
      \:\hat{T}_{N dir} (  E;\vec{p},\vec{k},\vec{q})\:
      \psi_\alpha(\vec{p})
\qc
\eea
where for the moment we assume the pion wave function to be a plane
wave. If we now make the local approximation as described 
at the end of Sec.~\ref{theonucl} by
putting the momentum of the incoming nucleon in the argument of
$\hat{T}_{N dir}$ equal to a constant $\vec{p}_o$, we can rewrite
\eq{amplexmom}:

\bea
\label{tloc}
T_{N dir} &\approx& Tr \left(
         \hat{T}_{N dir} ( E;\vec{p}_o,\vec{k},\vec{q}) \:\:
         \hat{\rho}_A(\vec{k}-\vec{q}) \right)
\qc
\eea
with the density matrix:

\bea
\label{nucten}
\hat{\rho}_A(\vec{p}) &=&  
\int  \frac{d^3 p'}{(2\pi)^3} \: 
\sum\limits_{\alpha \:occ.} \: 
      \psi_\alpha(\vec{p'}) \otimes
      \overline{ \psi}_\alpha(\vec{p'}+\vec{p}) \nn \\
 &=&
\int {d^3 x} \: \rme^{\rmi \vec{p} \vec{x}} \:
\sum\limits_{\alpha \:occ.} \: 
      \psi_\alpha(\vec{x}) \otimes
      \overline{ \psi}_\alpha(\vec{x})
=\int {d^3 x} \: \rme^{\rmi \vec{p} \vec{x}} \: \hat{\rho}_A(\vec{x})
\qp
\eea
{ The density matrix $\hat{\rho}_A(\vec{x})$ contains the complete
  information about the nuclear ground state.  The scalar, vector and
  tensor densities are obtained by taking $Tr(\Gamma \hat{\rho}_A)$
  with $\Gamma = 1, \gamma_o, \sigma^{0i}$, respectively. This is in
  contrast to the approach used in \cite{pieka} for the coherent
  photoproduction of $\eta$-mesons. There the production operator is
  rewritten using the free Dirac equation for in- and outgoing
  nucleons. As a consequence, the coherent photoproduction of
  $\eta$-mesons depends in \cite{pieka} only on the tensor density of
  the nuclear ground state.  }

One sees from Eqs.~(\ref{tloc}) and (\ref{nucten}), that in the local
approximation the differential cross section for coherent
photoproduction at a fixed energy involves the Fourier transformation
of a nuclear ground state density, i.e.~a nuclear form factor.
Although effects due to the non-locality of the production
operator and the distortions of the pion wave function by the nucleus
are superimposed,
the differential cross section is dominated by the
nuclear form factor. Hence the properties of the production operator
itself can only be studied when the energy dependence of the cross
section is considered.

\subsection{The nuclear wave function}
\label{nucwav}

It is clear from these considerations, that it is crucial to describe
the ground state properties of the nucleus realistically. The wave
functions in \eq{amplex} are taken from a relativistic
mean-field calculation using scalar and time-like vector potentials
$V_s$ and $V_v$, respectively:

\bea
(\bsl p\: - \: m \: -\: V_v \gamma_o \: -\: V_s) \psi_\alpha \:= \: 0
\qp
\eea
For the potentials $V^v $ and $V^s$ we assume a Woods-Saxon shape:

\bea
\label{nucpot}
V(r) \: = \: V_i^o\: \left( 1\: +\: 
      \rme^{\frac{(r - r_iA^{1/3})}{a_i }}\right)^{-1}
\quad ; i = v,s
\qp
\eea
The parameters we use for these potentials are given in Table
\ref{nucpotp}. They were determined such that the separation energies
\cite{nucdat1}, the root mean square radius of the charge density and
the charge form factors of $^{12} {\rm C}$ and $^{40} {\rm Ca}$
\cite{nucdat2} are well reproduced.  The resulting charge form factors
are shown in Figs.~\ref{ffc} and \ref{ffca}, together with the values
extracted from experiment \cite{nucdat2}.

{
  
  By using these wave functions, we neglect higher order correlations
  beyond mean field dynamics.  The question arises, to what extent the
  momentum structure of the nuclear ground state wave functions is
  adequately described by mean- field dynamics \cite{horo,furn}. The
  coherent production process leaves the nucleus in the ground state
  and probes, in the energy range considered here, primarily momenta
  inside the Fermi-sphere. Thus, a reliable description of nuclear
  form factors in a momentum range up to about twice the
  Fermi-momentum is most important. As seen in Figs.~\ref{ffc} and
  \ref{ffca}, the measured charge form factors are indeed well
  reproduced up to a momentum transfer of about 3 fm$^{-1}$, so that
  the relevant momentum range is obviously well described. Extended
  approaches (e.g.~\cite{corr,lenske}) are in fact showing that for
  spherical nuclei deviations from mean-field dynamics will become
  detectable only at rather high momenta corresponding to short-range
  processes which are not considered here.  A mean field version of
  the Walecka model has recently been used to treat the coherent
  photoproduction of $\eta$-mesons on nuclei \cite{pieka}.  }

\subsection{The pion-nucleus interaction}
\label{pinucint}

It is well known, that the coherent cross section depends strongly on
the interaction of the produced pion with the nucleus. 
This is taken into account by using
 a distorted pion wave function in \eq{amplex}, 
which is calculated from a position space optical
potential. An optical potential as given in \cite{stricker1,stricker2}
is used:

\bea
\label{pipot1}
2 E_\pi U_{opt}(r)&=& -4 \pi [ b(r)+B(r)] \nn \\
&& +4 \pi \grad \{ L(r) [c(r) + C(r)]\}\grad \nn \\
&& -4 \pi \left[ \frac{p_1-1}{2} \nabla^2 c(r) + 
                 \frac{p_2-1}{2} \nabla^2 C(r) \right]
\qc
\eea
where

\bea
b(r) &=& p_1 b_o \rho(r) \nn \\
c(r) &=& p_1^{-1} c_o \rho(r) \nn \\
B(r) &=& p_2 B_o \rho^2(r) \nn \\
C(r) &=& p_2^{-1} C_o \rho^2(r) \nn \\
L(r) &=& \left\{ 1 + \frac{4\pi}{3} \lambda[c(r) + C(r)]\right\}^{-1}
\nn 
\qc
\eea
with

\bea
p_1 = 1+\frac{E_\pi}{m_N}; \quad p_2 = 1+\frac{E_\pi}{2m_N}
\qp
\eea
$E_\pi$ is the total energy of the pion, $m_N$ is the mass of the
nucleon and $\rho$ is the nuclear density normalized to $A$, obtained
by summing the single particle densities in the potentials \eq{nucpot}
over the occupied states.  This optical potential contains $s$- and
$p$-wave interactions of the pion via $b(r)$ and $c(r)$ and the so
called true absorption via $B(r)$ and $C(r)$. The quantity $L(r)$
comes in because of the Lorentz-Lorenz-Ericson-Ericson effect
\cite{ew}.  The last term in \eq{pipot1} results from the so called
angle transformation \cite{ew}, which is a kinematical effect. The
parameter $\lambda$ is a real constant, while the quantities $b_o$,
$c_o$, $C_o$, and $C_o$ are complex and energy dependent.  In
\cite{stricker1,stricker2} these parameters are only given for pion
kinetic energies up to 50 MeV.  Since we want to compare our
calculations to the A2 data, we need an optical potential for pion
kinetic energies up to about 300 MeV. In order to have a consistent
parameterization over this range of energies, we adopt the same form
as in \eq{pipot1} also for higher energies.  The parameters were
determined by fitting the elastic scattering data of pions on $^{12}{\rm C}$
\cite{pidat}, including the nuclear Coulomb potential as done in
\cite{stricker1}.

Doing so one is confronted with the fact that there is a large
redundancy between some of the parameters, especially between $b_o$
and $B_o$ and between $c_o$ and $C_o$ \cite{seki}. Consequently, a
naive fit of all parameters leads to unphysical values. We
therefore put $B_o=0$ and $C_o=0$ and determine $b_o$ and $c_o$ as
complex, effective parameters.  In \fig{piel} we show the cross
sections for elastic pion scattering on $^{12}{\rm C}$ in comparison to the
experimental data. In order to have one more set of parameters at
higher energies, we included elastic pion scattering on $^{16}O$ at
$T_\pi$ = 330 MeV into the fitting procedure. For $T_\pi$ = 100 MeV
and $T_\pi$ = 157 MeV, we were not able to find a parameter set that
led to agreement with the data beyond the second minimum of the
differential cross section, without showing discrepancies at smaller
angles. In these two cases we chose the optical potential parameters
such that the data are reproduced well for angles up to the second
minimum. In order to obtain the optical parameters as smooth functions
of the pion energy, we have interpolated our fit results as shown in
\fig{pipotparm}.

{ We expect that the empirical potential thus obtained describes the
  final state interactions of the pion in the case of $^{12}{\rm C}$
  sufficiently well in an energy range from threshold up to the
  $\Delta$-resonance.  The structure of this potential is different
  from potentials that were obtained from microscopic calculations
  \cite{stricker1,garcia}, for which we also show results in
  \fig{piel}. We will compare the effects of our potential in the
  coherent photoproduction to the effects of potentials that were
  extracted from microscopic models in Sec.~\ref{dwia}. }

\section{Results}
\label{results}

\subsection{PWIA}

The model described in the previous section was used to calculate
differential and total cross sections for the coherent photoproduction
of pions on nuclei. In \fig{sigtotpwia} and \fig{dsdopwia} we show our
results for the plane wave impulse approximation (PWIA), i.e.  with a free
outgoing pion. As has been discussed at the end of \msc{theocoh}, the
total \cs shows the resonant character of the elementary production
operator, while the shape of the differential cross section results
from the nuclear form factor (multiplied with $\sin^2\theta$). In
\fig{sigtotpwia} we also show the contributions of the different
diagrams. Clearly, the $\Delta$-resonance dominates the process, but
the other diagrams lead to sizable corrections that should not be
neglected. Note the destructive interference between the Born terms
and the other graphs.
The omega contribution comes mainly from the vector coupling
of the omega (\eq{lint}), the tensor coupling contributes only
negligibly.

  In order to compare to the non-relativistic, local approach used in
  the previous DWIA calculations, we performed a local calculation using
  the non-relativistic reduction of the direct $\Delta$-graph as given
  in \cite{laget}, which has also been used in \cite{boffi1,chumba1}.
  The direct $\Delta$-graph accounts for most of the
  total cross section (see \fig{sigtotpwia}).  In order to compare
  this to our result for the direct $\Delta$-graph, we have used our
  set of parameters (\eq{couplcons}) including the energy dependent
  $\Delta$-width from \eq{deldec}. The struck nucleon was assigned an
  energy equal to the average energy of a bound nucleon in $^{12}{\rm
    C}$ (cf.~\eq{econs}).  The three-momentum $\vec{p}_i$ of the
  struck nucleon in the photon-nucleus cm-system was assumed to be
\bea
\vec{p}_i = - \frac{\vec{k}}{A} - \frac{A-1}{2A} (\vec{k}-\vec{q})
\qc
\eea
which is often used in local DWIA calculations (see
e.g.~\cite{chumba1}).  { The nuclear form factor was taken to be the
vector form factor resulting from our ground state calculation, which
corresponds to the charge form factor shown in \fig{ffc}, corrected
for the electromagnetic form factor of proton and neutron. 

The result of this calculation for the direct $\Delta$-graph is shown
as the dotted curve \fig{sgnrel} for $^{12}{\rm C}$ together with the
result of `exact' calculations (solid curve).  In
\cite{carrasco,boffi2} it is found that in a non-relativistic approach
there are additional corrections due to the transformation between the
photon-nucleus and the photon-nucleon cm-frame. In \cite{carrasco} it
is shown that the inclusion of this transformation corresponds to
multiplying the amplitude with a factor $m[(m+E_\gamma)\sqrt{1-v^2}]^{-1}$,
where $m$ is the mass of the nucleon and $v$ is the 
boost velocity from the photon-nucleus to the photon-nucleon cm-system.
Including this factor in our non-relativistic, local calculation leads
to the dashed curve in \fig{sgnrel}. It is an advantage of the
relativistic approach that these effects are correctly included
when the Lorentz invariant amplitude (\eq{amplex}) is calculated.  }

The difference between the the solid and the dashed curve in
\fig{sgnrel} cannot uniquely be separated into relativistic and
non-local effects.  In order to study the influence of the
relativistic nuclear structure we performed calculations where we
assumed the free relations between the upper and lower components of
the nuclear wave functions, which led only to a slight decrease of the
solid curve in \fig{sgnrel}.  The difference between the solid and the
dashed curves in \fig{sgnrel} is therefore a result of the
relativistic, non-local treatment of the production operator. In a
non-relativistic framework non-local effects have been shown to be
important in the photoproduction of charged pions in \cite{li,tiator}.

\subsection{DWIA}
\label{dwia}

We now take the final state interaction of the pion with the nucleus
into account by means of the optical potential described in Sec.~\ref{pinucint}.
The resulting total and differential cross sections for $^{12}{\rm C}$ and
$^{40}{\rm Ca}$ are shown in Figs.~\ref{sigtotdwia}, \ref{dsdodwiac} and
\ref{dsdodwiaca} in comparison to PWIA results. For the case of
$^{40}{\rm Ca}$ we took the same parameters for the pion optical potential
as for $^{12}{\rm C}$. For $E_\gamma \lapprox$ 230 MeV, i.e. for $T_\pi
\lapprox$ 100 MeV, where the absorption is relatively small (cf.
\fig{pipotparm}), the distortion of the pion wave function leads to an
increase of the total cross section, indicating the importance of
off-shell effects in the pion distorted wave.  For higher photon
energies the \cs is strongly reduced because of the large imaginary
part of the pion optical potential.  This reduction is stronger in the
case of $^{40}{\rm Ca}$, since the absorption of pions increases with the
nuclear mass. Besides that, the curves for the two different nuclei
are very similar in shape, and differ mainly by a global factor.
Making an ansatz $\sigma \sim A^\alpha$ we find $\alpha \sim$ 0.7-0.8 for
the case of PWIA.  This rather weak A-dependence results from the
different momentum dependence of the nuclear form factor for $^{12}{\rm C}$
and $^{40}{\rm Ca}$ (cf.  Figs.~\ref{ffc} and \ref{ffca}): For a given
momentum transfer, the $^{40}{\rm Ca}$ form factor is much smaller than the
one for $^{12}{\rm C}$, which compensates for the fact that there are $A$
amplitudes to be summed for a given nucleus (\eq{amplex}). For the
DWIA results we find $\alpha \sim$ 0.5-0.7,
depending on the photon energy, which reflects the pion absorption.

In \fig{a2} the result of a DWIA calculation for $^{12}{\rm C}$ is
compared to the A2 data.  The position of the maximum in the data is
well reproduced, but the height of this maximum is underestimated by
more than a factor of two.  Before we turn to improvements of the DWIA
calculation, we want to explore two possible sources of uncertainty.

First one has to be aware that two different but
phase-equivalent optical potentials might yield the same elastic
scattering cross section, but lead to a different behavior of the pion
wave function in the nuclear interior \cite{john}.  Since in \eq{amplex} 
the pion wave function within the nucleus, rather than its
asymptotic behavior, is relevant, two phase equivalent optical
potentials might in principle lead to different cross sections for the
coherent photoproduction of pions. 

{ In order to explore this ambiguity, we have also performed
  calculations using the optical potential of \cite{garcia} and of
  \cite{stricker1,stricker2}. As can be seen from \fig{piel}, these
  potentials yield pion elastic scattering cross sections that are,
in the respective energy ranges, 
  comparable to the ones obtained with our optical potential.
In \cite{garcia}
  the results of a microscopic $\Delta$-hole model calculation for the
  $\Delta$-resonance in nuclear matter \cite{oset2} have been used to
  construct a position-space optical potential for the pion.  In
  \fig{a2dh} we show the result of a calculation employing this optical
  potential as well as  our result in DWIA and PWIA.
Both potentials lead
  to a rather similar behavior of the cross section, especially at
  higher photon energies. For the total cross sections, the agreement
  between calculations employing these two different optical potentials
  is qualitatively the same as for the differential cross section in
  \fig{a2dh}.}  We have also performed calculations with the parameter sets
given in \cite{stricker1} and \cite{stricker2} for $T_\pi$ = 50 MeV
($E_\gamma$ = 184 MeV). Using these parameter sets, in which the
parameters $B_o$ and $C_o$ are non-zero, we get a total cross section
that differs from the one obtained with our pion optical potential by
less than 10 \% at this specific pion energy.  { We thus conclude,
  that our empirical parameterization for the pion optical potential
  accounts for the essential features of the pion nucleus interaction
in the case of $^{12}{\rm C}$ and  $^{40}{\rm Ca}$.
}

As mentioned above, the $\omega NN$ vector coupling constant and the
form factor (\eq{omff}), which lead to the $\omega$ contribution in
\fig{sigtotpwia}, are not accurately known.  In the elementary
photoproduction of neutral pions the contribution of the omega graph
to the differential cross section is about 10 \%, depending on the
angle.  To show the dependence of our result for the coherent
photoproduction on the omega parameters, we show in \fig{a2slo}
calculations employing different values of $g_{\omega NN}^v$ and
$\Lambda$.  The values $g_{\omega NN}^v=17$ and $g_{\omega NN}^v= 8$
represent the upper and lower bound of possible values for this
coupling. The resulting uncertainty in our results is much smaller
than the discrepancy between our DWIA results and the A2 data.

\subsection{Medium effects}
\label{mediumeffects}

{ By using an optical potential to determine the distorted wave
  function of the pion, we already include the effect of the
  interaction of the produced pion with the nucleus. The intermediate
  particles contributing to the production operator, however, also
  interact with the nuclear medium.}

For the nucleon propagator we include the resulting medium effects by
replacing the free propagator with the one obtained from the
mean-field in which the bound state wave functions were calculated. We
thus employ a dressed nucleon propagator $G_N$ that fulfills

\bea
\label{nucpropim}
(\bsl p\: - \: m \: -\: V_v \gamma_o \: -\: V_s)
  G_N(p_o;\vec{x},\vec{y})  \:= \: \rmi \delta^3(\vec{x}-\vec{y})
\qp
\eea
Since $G_N$ is calculated in the mean-field approximation, it 
contains the static properties of a struck nucleon
interacting with the residual nucleus.  
 Although it can certainly not account for the full dynamics of
  the struck nucleon interacting with the residual nucleus, 
it is the
natural extension of the elementary model to reactions on the nucleus,
in which the bound and the intermediate nucleon are treated
consistently.  In \fig{sigtotmeff} we show results of calculations
for $^{12}{\rm C}$ using the free as well as the dressed nucleon propagator
from \eq{nucpropim}.  The increase of the cross section results from the
fact, that the total contribution of the nucleon terms is the result
of an approximate cancellation of direct and exchange graph. In the
case of the dressed nucleon propagator, this cancellation is more
complete, leading to a smaller net contribution. Because of the
destructive interference of the Born terms with the other
contributions, this leads to a larger cross section.

 For the $\Delta$-resonance we can use the $\Delta$-hole model as
  a guideline. In \cite{koch,koch0} it is shown, that the medium
  modifications of the $\Delta$, in addition to a distortion of the
  pion, have a clear effect on the coherent cross section.  It is well
known, that the $\Delta$ feels an attractive potential of about -30
MeV at normal nuclear density \cite{ew}, which has an effect on any
production process involving the $\Delta$.  Indeed, in $(^3{\rm He},t)$ and
$(d,^2\!{\rm He})$ reactions a shift of the $\Delta$-peak relative to the
reaction on the proton has been observed \cite{ellegaard,contardo}.  A
considerable part of this shift is, however, due to trivial effects
e.g. the energy dependence of the nuclear form factor, so that it is
only partially a consequence of in-medium modifications of the
$\Delta$ \cite{delorme,udegawa}. In contrast to that, the
$\Delta$-peak remains essentially unshifted in the photoabsorption on
nuclei \cite{bianchi}.

Calculating the full, relativistic $\Delta$-propagator in the presence
of nuclear potentials is out of the scope of this work. We take the
attractive self-energy of a $\Delta$ within a nucleus schematically
into account by shifting its mass in the production operator. In
\fig{sigtotmeff} we also show the result of a calculation with the
$\Delta$-mass reduced by 30 MeV (full line).  A mass shift of the
$\Delta$ increases the \cs at lower energies and decreases it at
higher energies. The value $\delta m_\Delta = -$30 MeV has been chosen
such that the absolute height of the maximum of the A2 data
is reproduced (\fig{a2meff}).  Since this number can only be
interpreted as an average over the entire nucleus, it seems somewhat
large in comparison to the depth of the $\Delta$-potential of 30 MeV
\cite{ew}.  It must at this point also be kept in mind, that the
actual value of $\delta m_\Delta $ needed to fit the data depends to
some extent on the value used for of the $\omega NN$ coupling
constant.

Besides the mass, the width of the $\Delta$ is also modified in the
medium. A simple analysis taking into account Pauli blocking and
collisional broadening leads to a width that is rather independent of
the nuclear density \cite{effe}, but delta-hole calculations yield a
$\Delta$ self-energy, which corresponds to an increase of the width by
as much as 60 MeV \cite{oset2}.  { Including this effectively by
  adding a constant shift of 30 MeV to the energy dependent
  $\Delta$-width in \eq{deldec} changes the result depicted in
  \fig{a2meff} only at energies beyond the maximum, where the cross
  section is rather small.   Thus for the A2
data we can clearly distinguish between the effect of a modified mass
and a modified width of the $\Delta$.

{ This procedure, to take into account the distortions of the
  produced pion via an optical potential and then include medium
  modifications of the intermediate $\Delta$ by changing its mass and
  its width independently, 
mimics the essentials of the $\Delta$-hole model. 
The qualitative agreement with the
  $\Delta$-hole model can also be seen in \cite{koch}, where the
  amplitude for the coherent photoproduction of pions in the
  $\Delta$-hole model is rewritten in such a way, that it contains a
  distorted pion wave function and a modified $\Delta$-propagator.  In
  this study it is found, that the modifications of the
  $\Delta$-propagator, in addition to the pion distortions, have
  considerable effects.  }

As can be seen in \fig{a2meff}, the inclusion of a dressed nucleon
propagator and a lowered $\Delta$-mass leads to  agreement of our 
calculation
with the data around the maximum. At higher energies, the situation is
not clear. The in-medium \po seems to lead to an underestimation of
the data for $E_\gamma>$ 350 MeV. More experimental data, especially
at smaller angles, are needed in this energy range before definite
conclusions can be drawn.

Since there is no reliable information available on the properties
of an $\omega$-meson at space-like momenta in nuclear matter, we did
not consider in-medium modifications of the $\omega$-graph.

The special energy dependence of the differential cross section at a
constant angle is the result of an interplay between the energy
dependence of the \po and the nuclear form factor. In order to
circumvent the energy dependence of the form factor, and to eliminate
the trivial $\sin^2 \theta$ term (\eq{sin2}), we show in \fig{q} the
differential \cs for a constant momentum transfer $|\vec{q}|$ = 0.1
GeV, divided by $\sin^2 \theta$ as a function of the photon energy for
$^{12}{\rm C}$ and $^{40}{\rm Ca}$.  When plotting the cross section
this way, the effects of in-medium modifications of the \po are
magnified and become visible over a large range of energies.
Especially an increased $\Delta$-width now leads to a pronounced
effect.  Thus the new TAPS data, that will cover a large range of
angles for each energy \cite{krusche}, will provide an opportunity to
test our method of treating medium effects further.

{ In order to also compare to calculations using the $\Delta$-hole
  model, we show in \fig{sgimdo} the total cross section for
  $^{12}{\rm C}$ using a medium modified production operator taking
  all graphs into account, as well as that obtained by only including
  the direct $\Delta$-term.  The result for the direct $\Delta$ term
  agrees qualitatively with results of $\Delta$-hole calculations. As
  compared to the most recent publications using the $\Delta$-hole
  model \cite{carrasco,lakti}, the height of the maximum in the cross
  section is in our model about the same as in \cite{lakti}, but
  somewhat above the result of \cite{carrasco}.  The position of the
  maximum is, however, around $E_\gamma$=220 MeV in
  \cite{carrasco,lakti}, compared to 260 MeV in our model.  In
  \cite{koch,saha} the $\Delta$-hole model is used in combination with
  an empirical multipole decomposition of the production operator,
  thus including non-resonant contributions. Our total result in
  \fig{sgimdo} agrees with the result of these studies with respect to
  the position of the maximum, but is about 15\% higher.  }

\section{Summary and  Conclusions}
\label{sum} 

We have presented a relativistic and non-local model for the coherent
photoproduction of pions on { spin-zero} nuclei, { which is
  applicable from threshold up to the $\Delta$-region.}
Since the
calculation for this process is gauge invariant, and since only a
minimum amount of theoretical input is needed, this reaction
represents an attractive field to test our understandings of hadrons
and nuclei.

The production operator was derived from an effective Lagrangian for
which the free parameters were fixed by comparison to the data for
elementary processes. Applying this production operator to the
photoproduction on a nucleus involves an off-shell extrapolation with
respect to the nucleon. In our model we used the off-shell behavior
following naturally from an effective field theory.  The validity of
this extrapolation, can only be tested in comparison to experimental
data.  { Employing a mean field model for the nuclear wave function we
  neglected higher order effects from many-body correlations.}

Our results depend somewhat on the value used for the $\omega NN$
coupling constant, which is uncertain by about a factor of two.
Employing the same \po as in vacuum leads to a good reproduction of
the shape of the A2 data, but underestimates the absolute value by
about a factor of two.  Agreement with these data could be achieved by
taking into account medium modifications of the \po via modifications
of the nucleon- and $\Delta$-propagators.

We conclude that the relativistic DWIA, employing a realistic
production operator is an adequate approach to the coherent
photoproduction of pions on nuclei.  More experimental data on this
process will present a further test of our model and will allow more
conclusions about the role of medium effects within our model.

\acknowledgments
We would like to thank the A2-collaboration at MAMI for providing us
with their data prior to publication.
We gratefully acknowledge many helpful discussions with R. Shyam 
in an early stage of this work. 
We also would like to thank Thomas Feuster for performing the
elementary cross section calculations.

\begin{table}
\begin{tabular}{|c|c|c|c|c|c|c|} 
 Nucleus & $V _v$ & $ r_v $ & $ a_v $ & $V_s$ & $ r_s $ &
$a_s$  \\
     &(\footnotesize{MeV}) &
 (\footnotesize{fm}) & (\footnotesize{fm} ) & (\footnotesize{MeV}) &
  (\footnotesize{fm}) & (\footnotesize{fm}) \\ \hline
$^{12}C $  & 385.7 & 1.056 & 0.427 & -470.4 & 1.056 & 0.447 \\ \hline
$^{40}Ca $ & 348.1 & 1.149 & 0.476 & -424.5 & 1.149 & 0.506 \\ 
\end{tabular} 
\caption { 
Strengths, reduced radii and diffusivities for the relativistic scalar 
and vector mean-field potententials, respectively.
}
\label{nucpotp}
\end{table}

\renewcommand{\baselinestretch}{1.0}
\begin{figure}[!ht]
  \centerline{ \includegraphics[width=8cm]{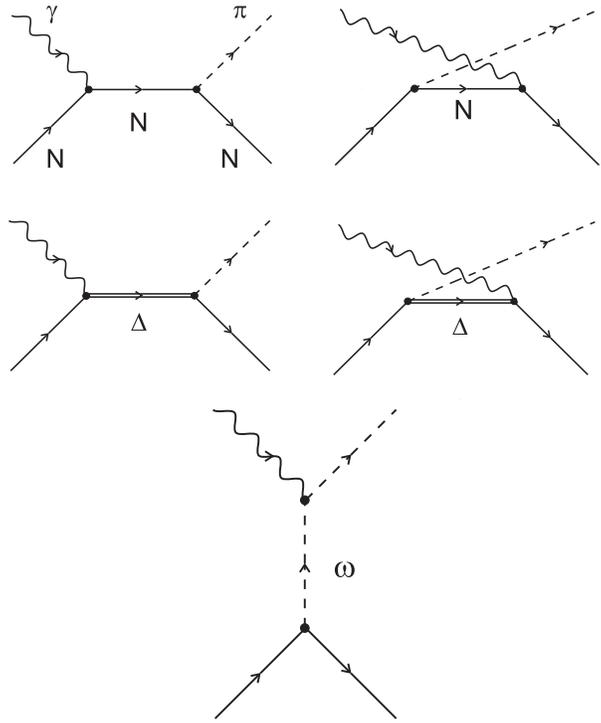} } 
  \caption{Feynman diagrams contributing to the photoproduction 
    of neutral pions.}
\label{diagrams} 
\end{figure} 

\begin{figure}[!ht]
  \centerline{ \includegraphics[width=13cm]{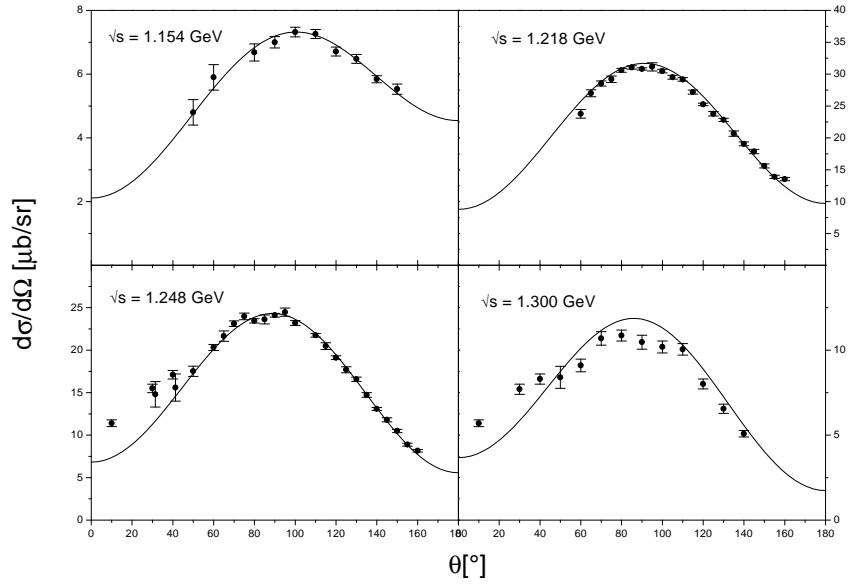} }
  \caption{
    Differential cross section for $\gamma p \to p \pi^o$ for four
    different cm-energies $\sqrt{s}$.  The data are taken from
    \protect \cite{pdat}.  }
\label{gamnuc} 
\end{figure} 

\begin{figure}[!ht]
  \centerline{ \includegraphics[width=13cm]{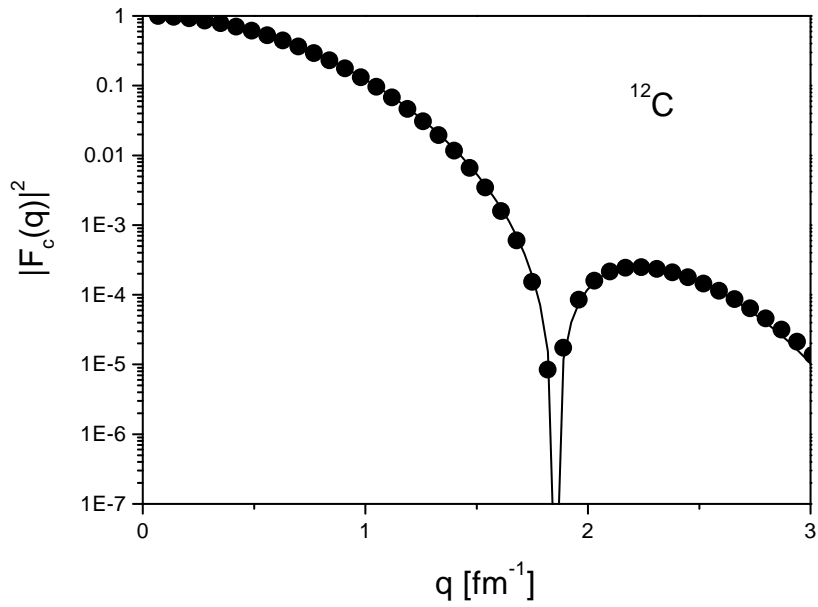} } 
  \caption{Charge form factor for $^{12}{\rm C}$ resulting from the wave
    functions used in this work (solid line) in comparison to values extracted
    from experiment \protect \cite{nucdat2} (full circles).}
\label{ffc} 
\end{figure} 

\begin{figure}[!ht]
  \centerline{ \includegraphics[width=13cm]{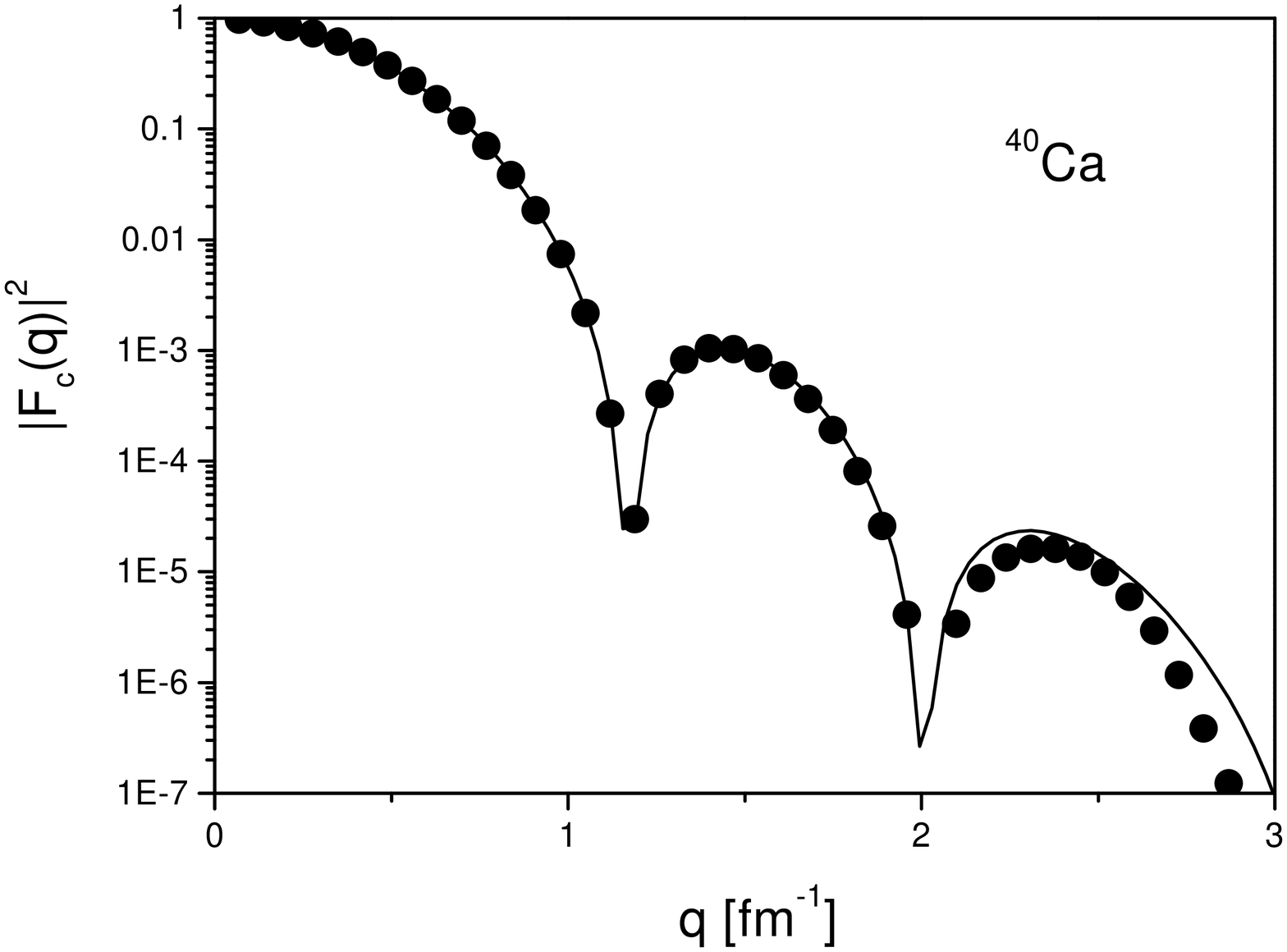} } 
  \caption{Same as \protect \fig{ffc}, but for $^{40}{\rm Ca}$.}
\label{ffca} 
\end{figure} 

\begin{figure}[!ht]
  \centerline{ \includegraphics[width=14cm]{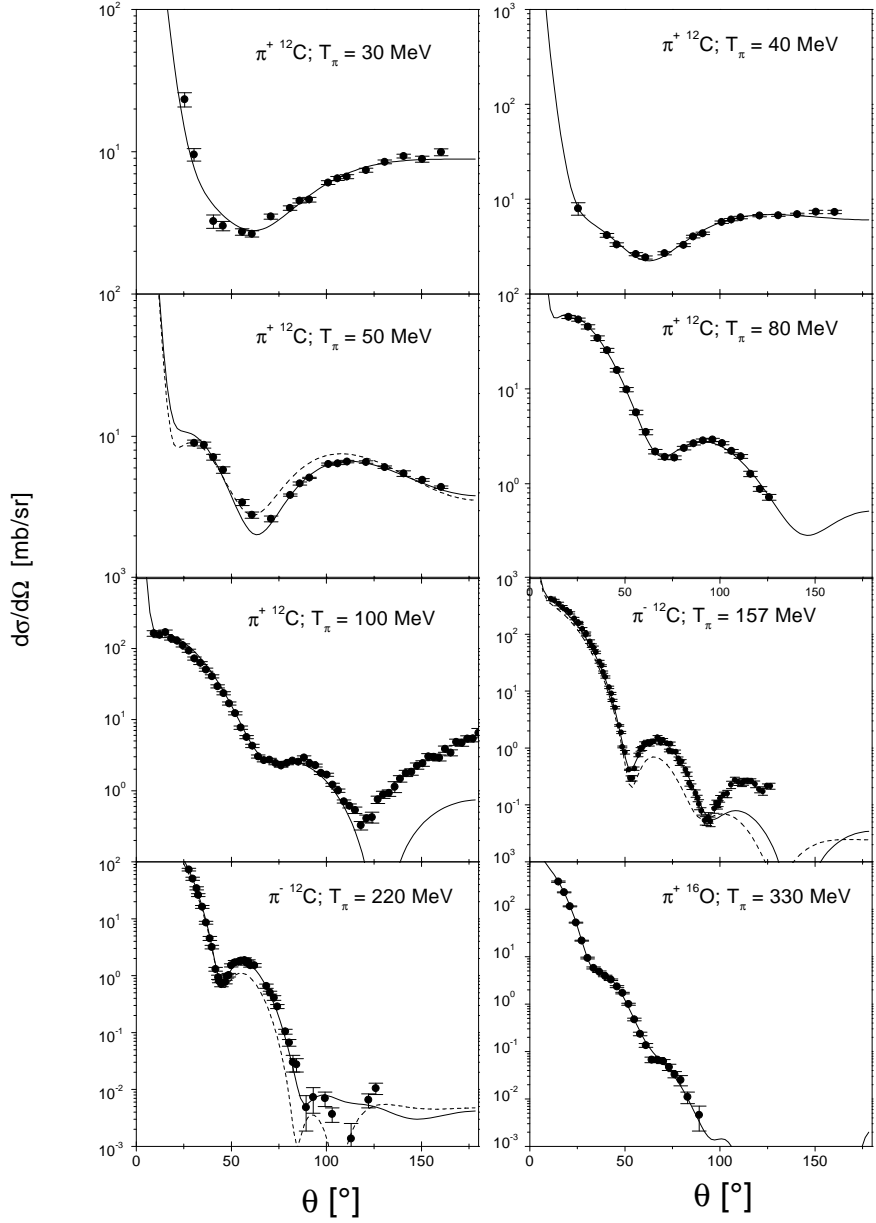} } 
  \caption{
    Fit results for elastic pion scattering as described in the
    text. The data are taken from Ref.
    \protect \cite{pidat}. The dashed lines show results
    for the parameter set $C$ from \protect \cite{stricker1} at 50 MeV
    and for the optical potential from \protect \cite{garcia} at 157
    and at 220 MeV.
\label{piel} } 
\end{figure} 

\begin{figure}[!ht]
  \centerline{ \includegraphics[width=13cm]{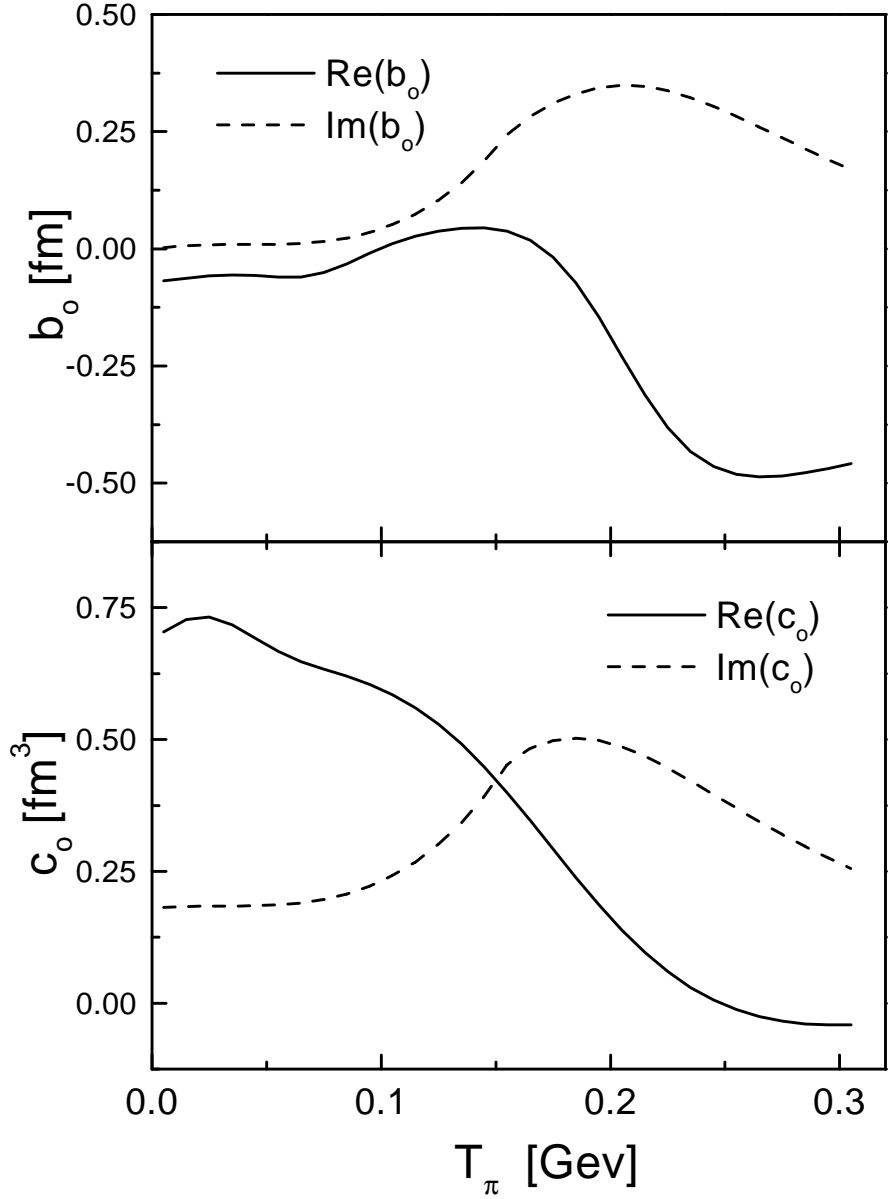} } 
  \caption{Energy dependence of the $s$- and $p$-wave 
parameters for the pionic optical potential resulting from
    our fit as described in the text.}
\label{pipotparm} 
\end{figure} 

\begin{figure}[!ht]
  \centerline{ \includegraphics[width=13cm]{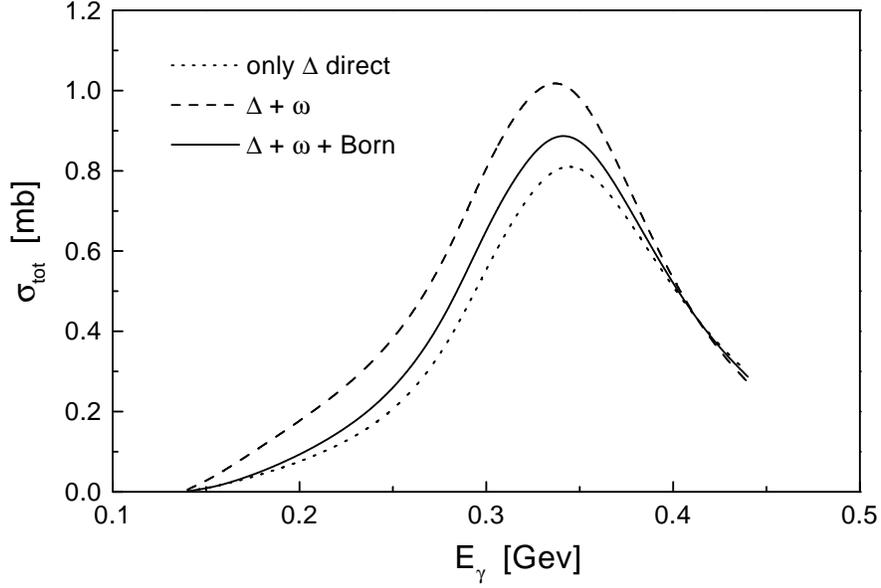} } 
  \caption{Contributions of the different fundamental diagrams to the 
    total cross section for 
    $^{12}{\rm C}(\gamma,\pi^o)^{12}{\rm C}$ in PWIA.}
\label{sigtotpwia} 
\end{figure} 

\begin{figure}[!ht]
  \centerline{ \includegraphics[width=13cm]{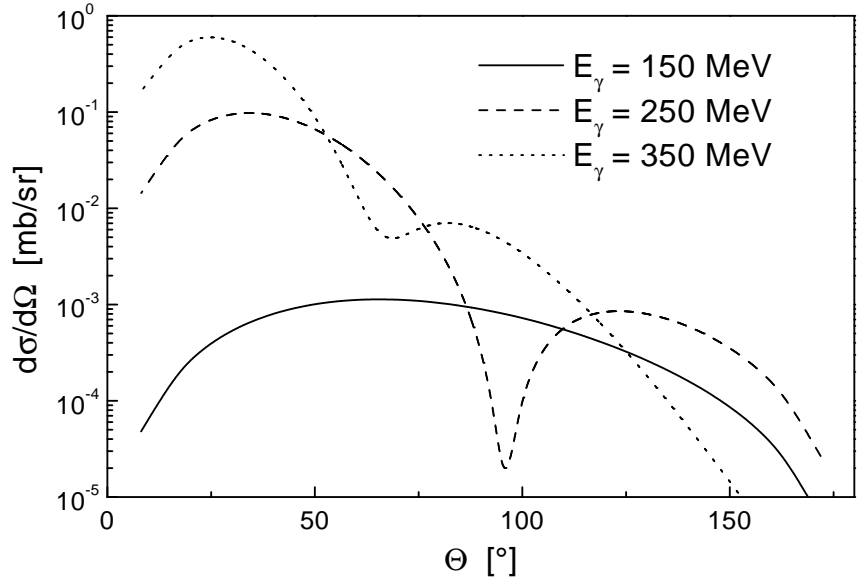} } 
  \caption{Differential cross section for 
    $^{12}{\rm C}(\gamma,\pi^o)^{12}{\rm C}$ in PWIA
for three different energies.}
\label{dsdopwia} 
\end{figure} 

\begin{figure}[!ht]
  \centerline{ \includegraphics[width=13cm]{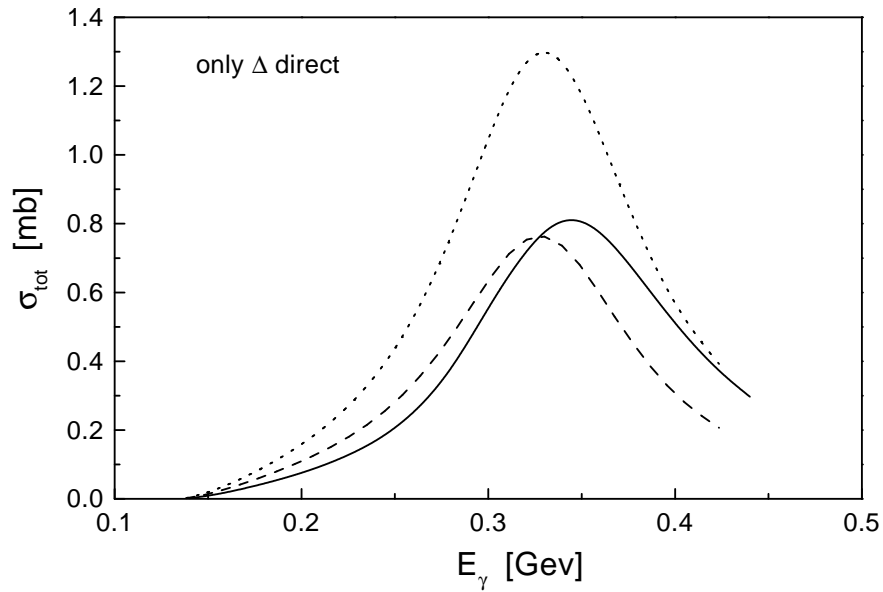} } 
  \caption{Total cross section for 
    $^{12}{\rm C}(\gamma,\pi^o)^{12}{\rm C}$ in PWIA resulting from
    the direct $\Delta$-graph (solid line) in comparison to
    non-relativistic, local calculations as in
    \protect\cite{boffi1,chumba1} (dotted line) and including the
    kinematical correction from \protect\cite{carrasco} (dashed
    line) as described in the text.
}
\label{sgnrel} 
\end{figure} 

\begin{figure}[!ht]
  \centerline{ \includegraphics[width=13cm]{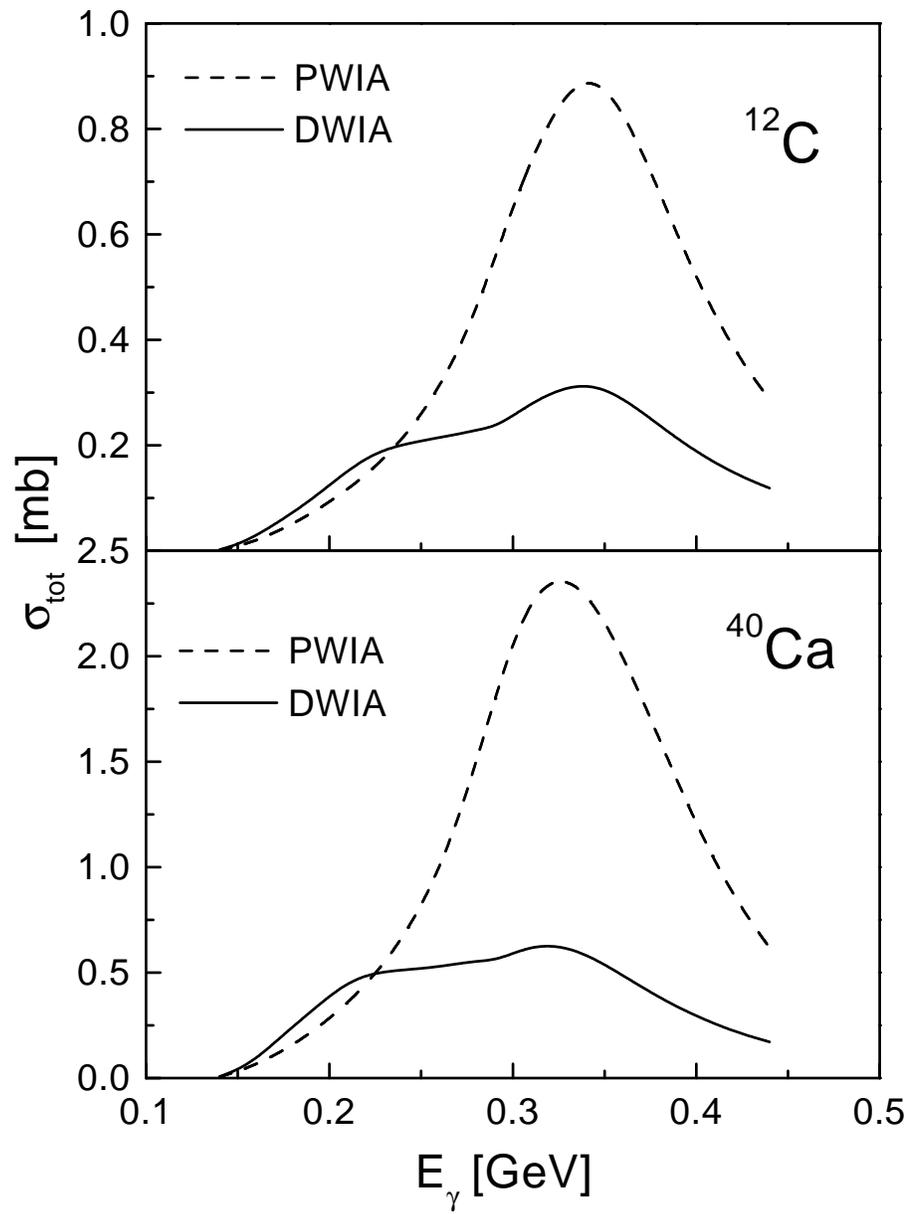} } 
  \caption{Total cross section in DWIA and PWIA for 
    $^{12}{\rm C}(\gamma,\pi^o)^{12}{\rm C}$ and
    $^{40}{\rm Ca}(\gamma,\pi^o)^{40}{\rm Ca}$.}
\label{sigtotdwia} 
\end{figure} 

\begin{figure}[!ht]
  \centerline{ \includegraphics[width=13cm]{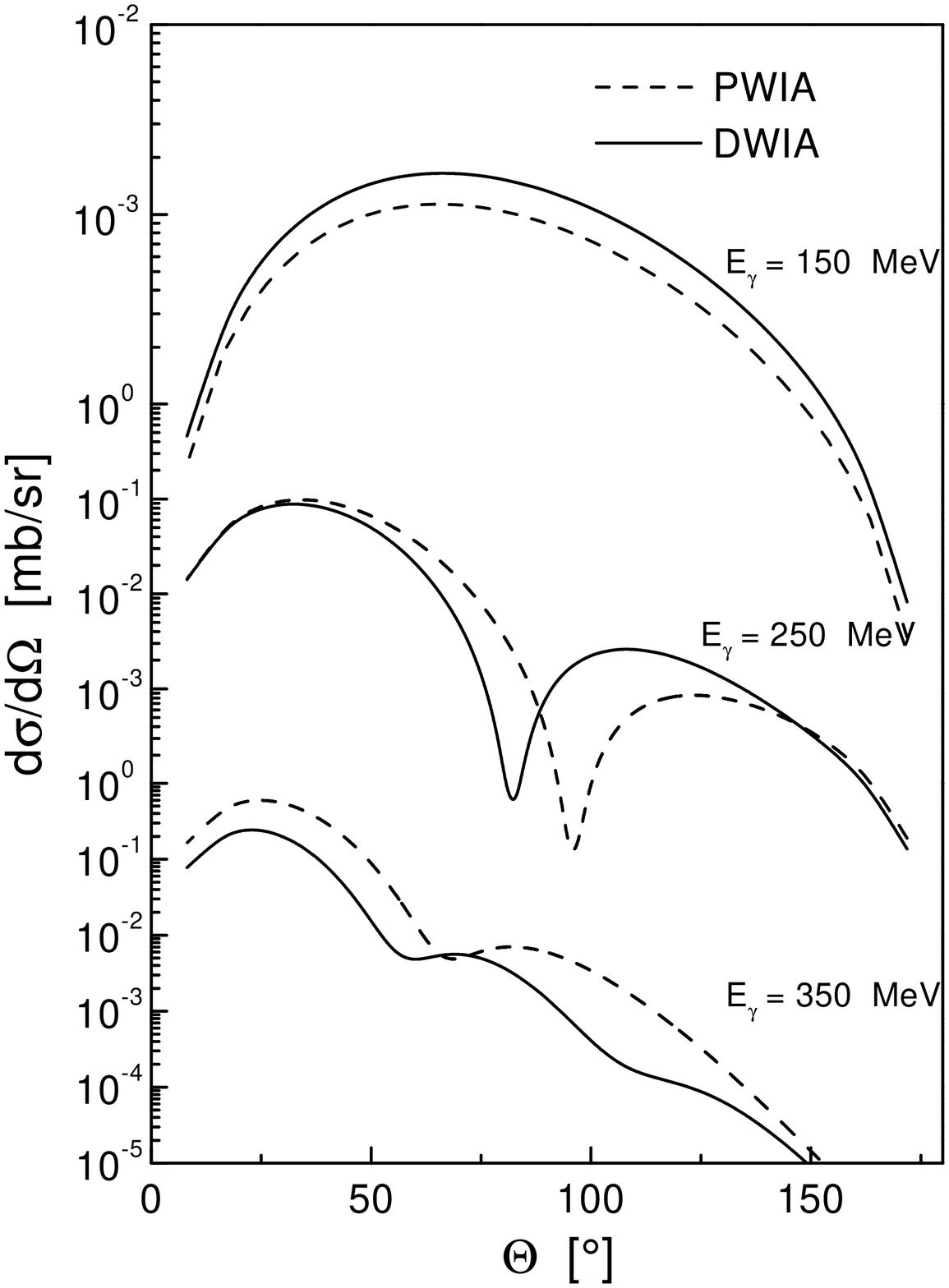} }
  \caption{Differential cross section in DWIA and PWIA for 
    $^{12}{\rm C}(\gamma,\pi^o)^{12}{\rm C}$.}  
\label{dsdodwiac} 
\end{figure} 

\begin{figure}[!ht]
  \centerline{ \includegraphics[width=13cm]{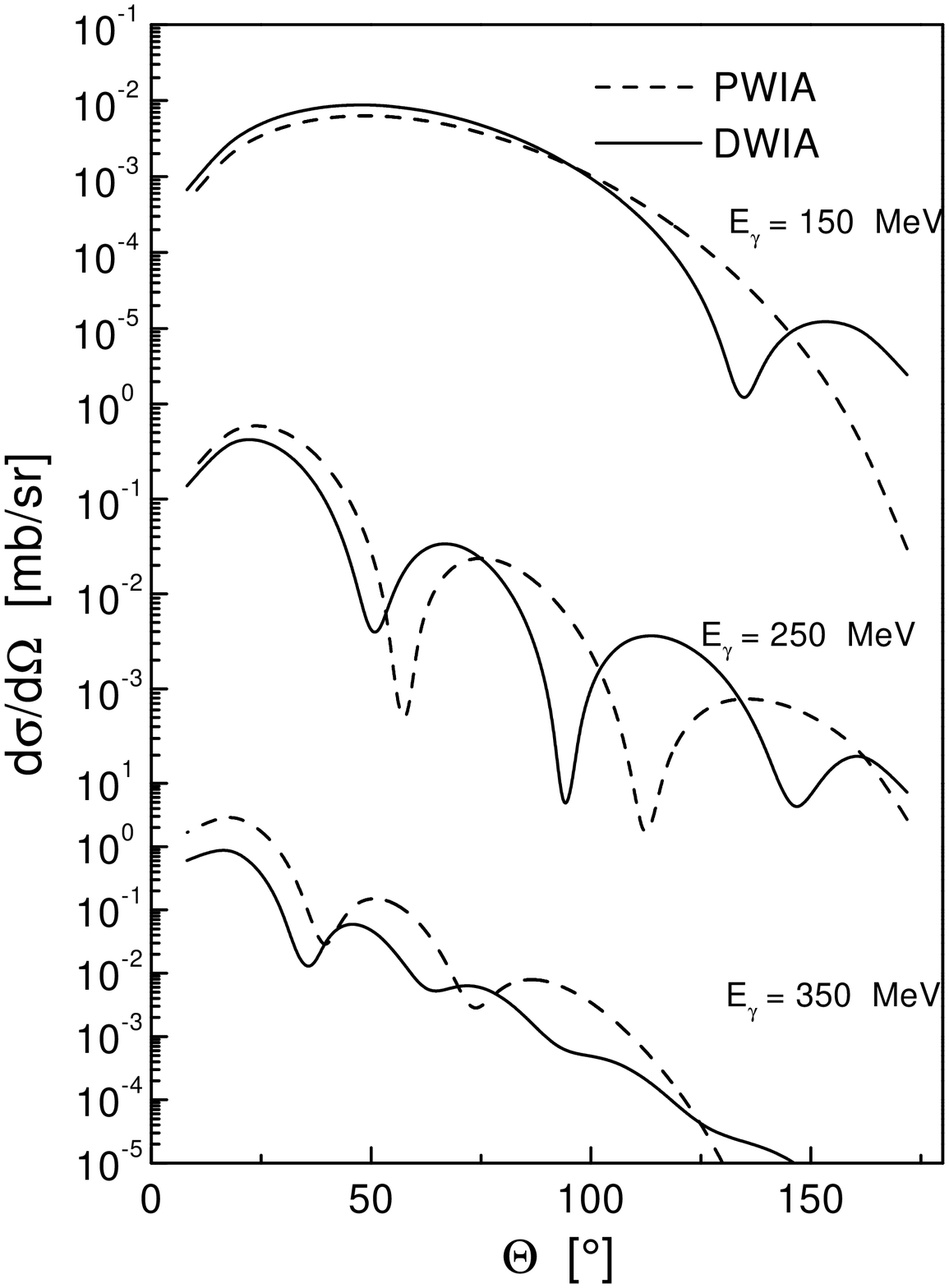} }
  \caption{Differential cross section in DWIA and PWIA for 
    $^{40}{\rm Ca}(\gamma,\pi^o)^{40}{\rm Ca}$.}  
\label{dsdodwiaca} 
\end{figure}

\begin{figure}[!ht]
  \centerline{ \includegraphics[width=13cm]{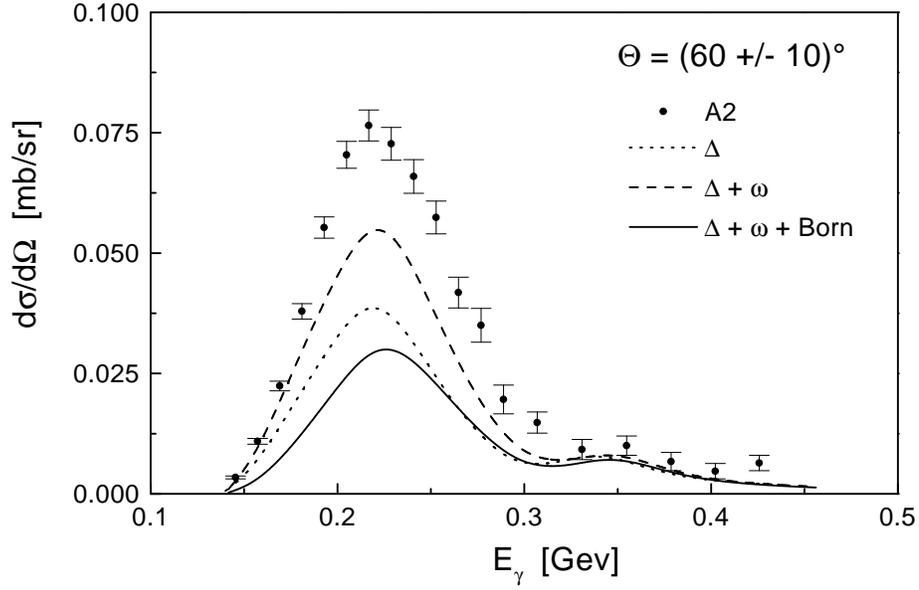} } 
  \caption{Comparison of the A2-data for $^{12}{\rm C}$
    to a DWIA calculation.}
\label{a2} 
\end{figure} 

\begin{figure}[!ht]
  \centerline{ \includegraphics[width=13cm]{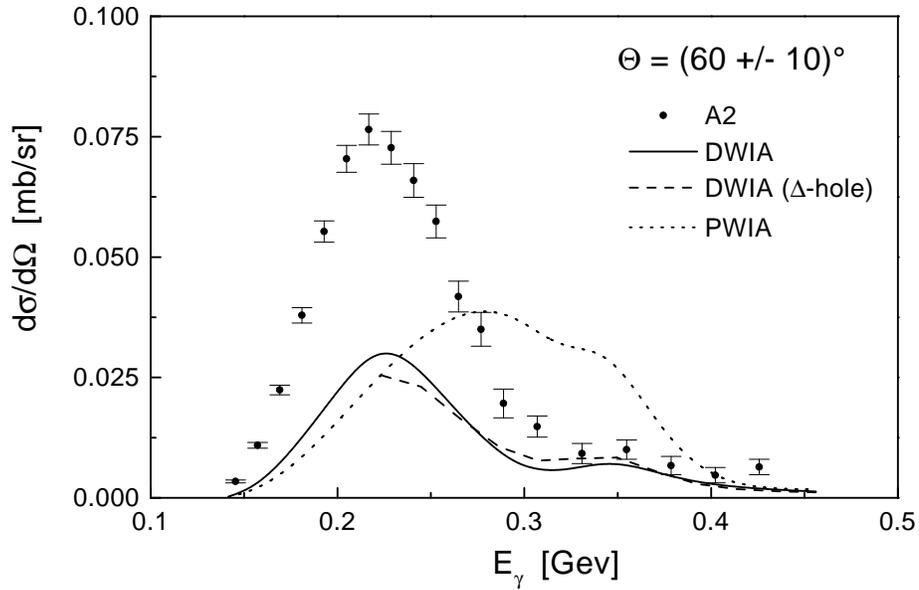} } 
  \caption{Comparison of the A2-data for $^{12}{\rm C}$
    to DWIA calculations employing two different optical potentials and
    a PWIA calculation.}
\label{a2dh} 
\end{figure} 

\begin{figure}[!ht]
  \centerline{ \includegraphics[width=13cm]{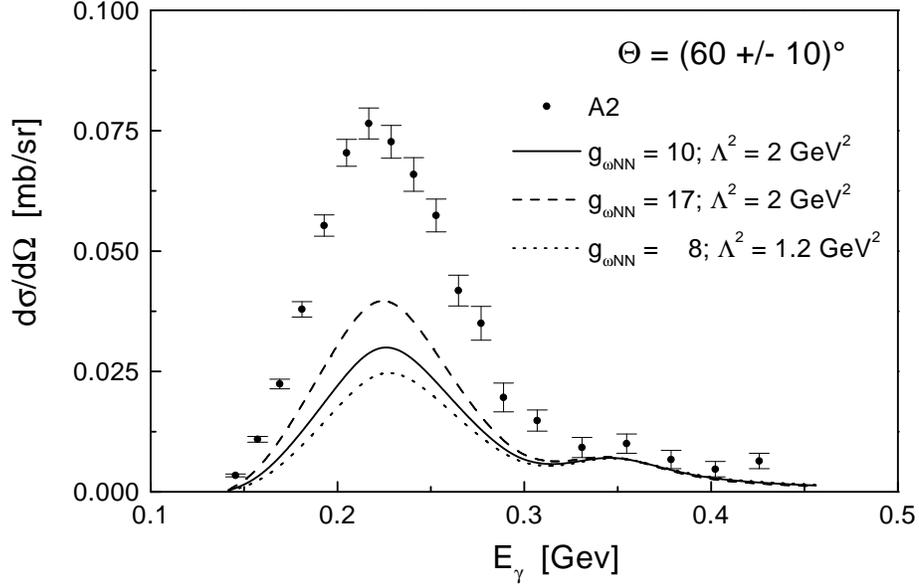} } 
  \caption{Comparison of the A2-data for $^{12}{\rm C}$
    to DWIA calculations employing different $\omega NN$ couplings and
    cutoffs. The
    solid curve was obtained with the values used throughout this work.}
  \label{a2slo} 
\end{figure} 

\begin{figure}[!ht]
  \centerline{ \includegraphics[width=13cm]{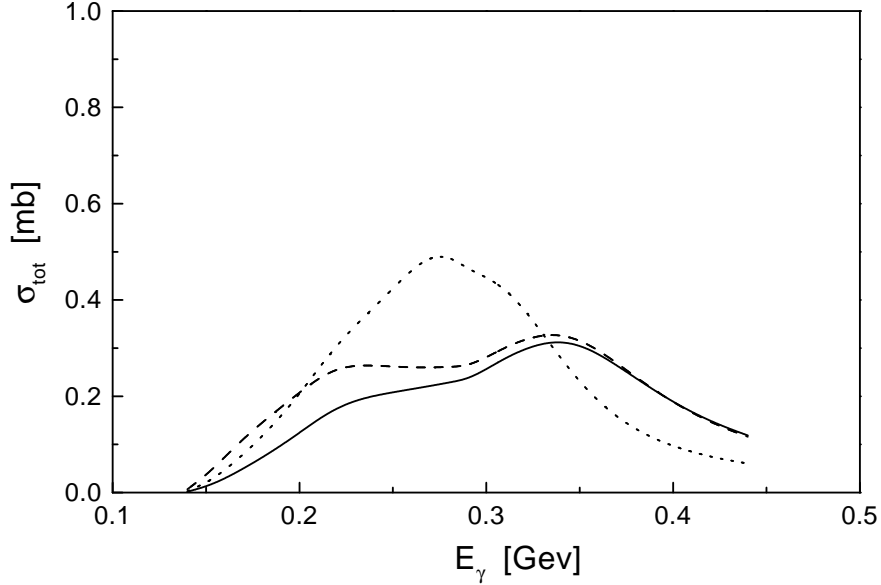} } 
 \caption{Total DWIA cross section for $^{12}{\rm C}(\gamma,\pi^o)^{12}{\rm C}$ 
   using the free
   production operator (full line), using the dressed nucleon
   propagator (dashed line) as described in the text, and with a mass
   shift of the $\Delta$ $\delta m_\Delta$ = -30 MeV (dotted line).}
  \label{sigtotmeff} 
\end{figure} 

\begin{figure}[!ht]
  \centerline{ \includegraphics[width=13cm]{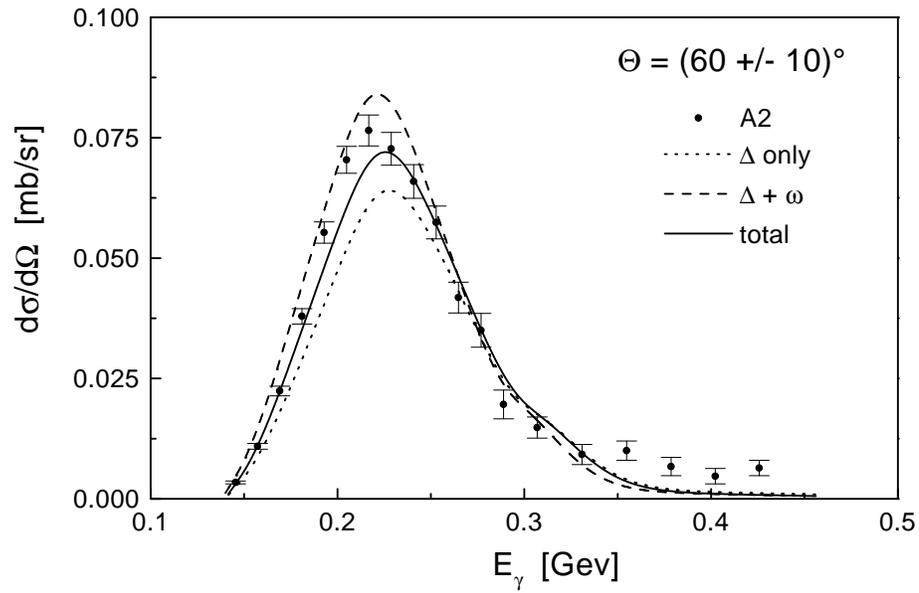} } 
 \caption{Comparison of the A2-data for $^{12}{\rm C}$
    to a DWIA calculation employing a medium-modified production
    operator with a dressed nucleon propagator and a shifted
    $\Delta$-mass ($\delta m_\Delta$=-30 MeV).}
  \label{a2meff} 
\end{figure}  

\begin{figure}[!ht]
  \centerline{ \includegraphics[width=13cm]{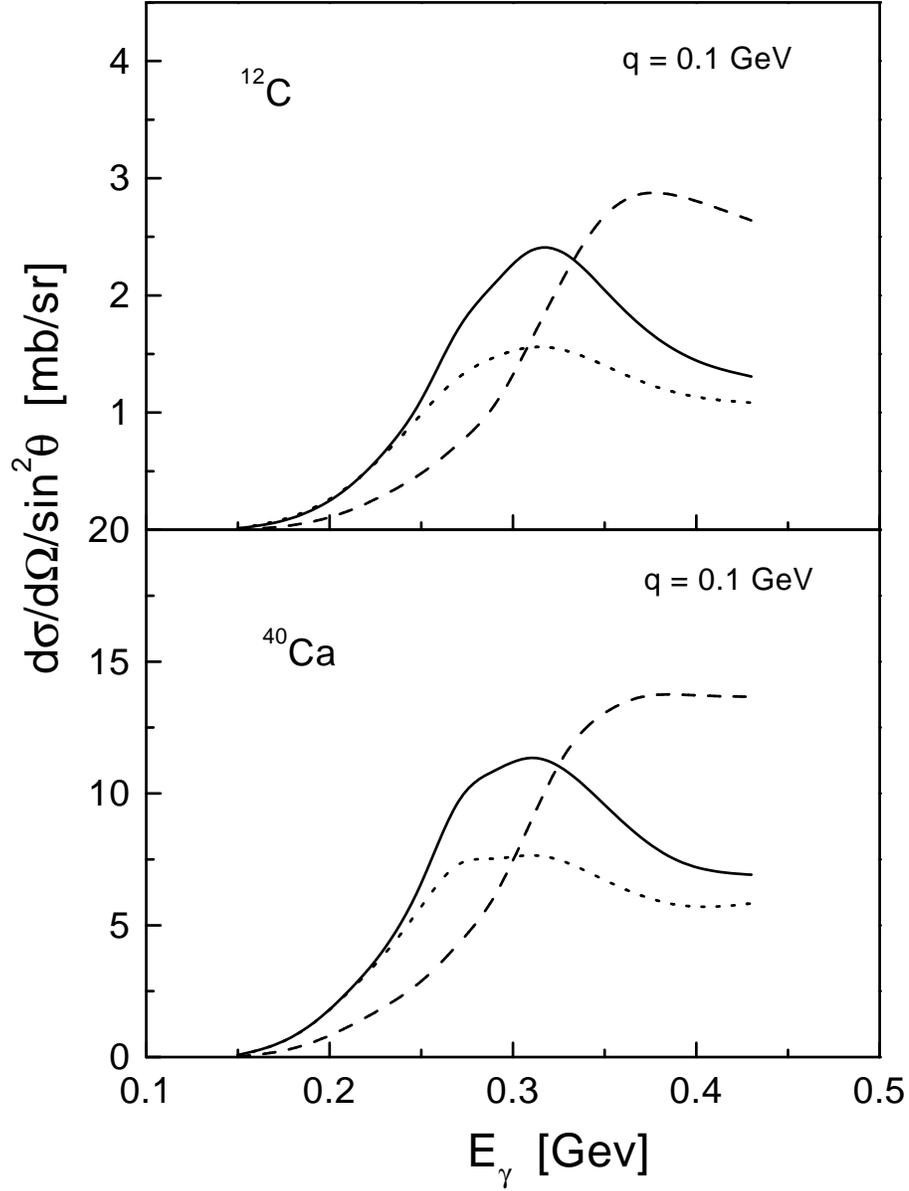} } 
 \caption{Differential cross section divided by $\sin^2\theta$ for 
   $^{12}{\rm C}(\gamma,\pi^o)^{12}{\rm C}$ and
   $^{40}{\rm Ca}(\gamma,\pi^o)^{40}{\rm Ca}$ for a fixed momentum transfer
   of 0.1 GeV: free production operator (dashed line), in-medium
   production operator with a dressed nucleon propagator and a
   $\Delta$-mass shift $\delta m_\Delta$=-30 MeV (full line), and with
   an in-medium production operator including additionally an
   increased $\Delta$-width $\delta \Gamma_\Delta$ = 30 MeV (dotted
   line).}
  \label{q} 
\end{figure} 

\begin{figure}[!ht]
  \centerline{ \includegraphics[width=13cm]{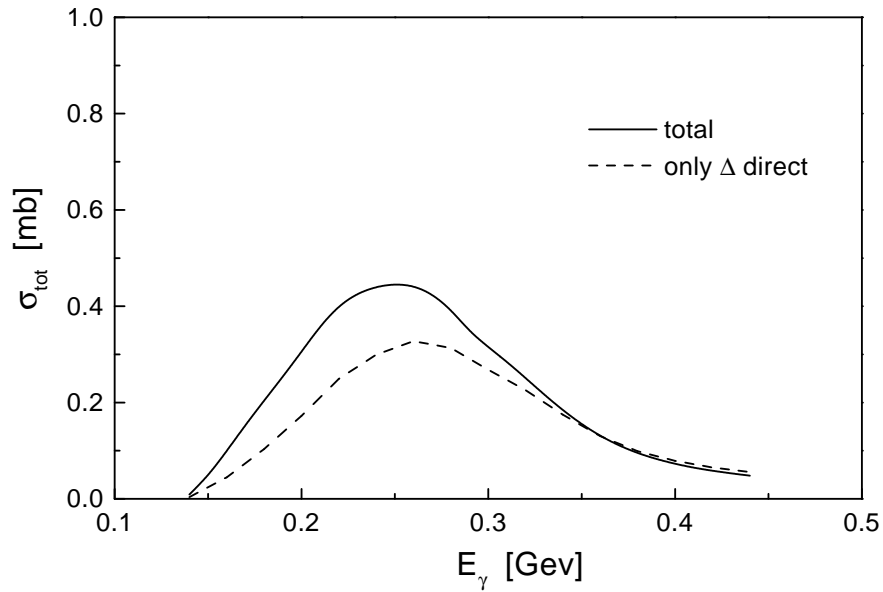} } 
 \caption{
   Total cross section for $^{12}{\rm C}(\gamma,\pi^o)^{12}{\rm C}$ using an
   in-medium production operator  ($\delta m_\Delta$=-30 MeV, $\delta
   \Gamma_\Delta$ = 30 MeV), for a complete calculation, as well as
   for a calculation where only the direct $\Delta$-graph is included. }
  \label{sgimdo} 
\end{figure}

\end{document}